\documentclass[sigconf]{acmart}

\AtBeginDocument{%
  \providecommand\BibTeX{{%
    \normalfont B\kern-0.5em{\scshape i\kern-0.25em b}\kern-0.8em\TeX}}}

\copyrightyear{2020}
\acmYear{2020}
\setcopyright{acmcopyright}
\acmConference[SIGIR '20] {43rd International ACM SIGIR Conference on Research and Development in Information Retrieval}{July 25--30, 2020}{Virtual Event, China}
\acmBooktitle{43rd International ACM SIGIR Conference on Research and Development in Information Retrieval (SIGIR '20), July 25--30, 2020, Virtual Event, China}
\acmPrice{15.00}

\acmDOI{10.1145/3397271.3401200}
\acmISBN{978-1-4503-8016-4/20/07}

\usepackage{makecell}

\usepackage{natbib}
\usepackage{graphicx}
\usepackage{fancyhdr}
\usepackage[ruled,vlined]{algorithm2e}
\include{pythonlisting}

\usepackage{color}
\newcommand{\grace}[1]{{\bf \color{red} [[Grace says ``#1'']]}}
\newcommand{\zhiwen}[1]{{\bf \color{blue} [[Zhiwen says ``#1'']]}}

\begin{document}
\fancyhead{}

\title{Balancing Reinforcement Learning Training Experiences in Interactive Information Retrieval}

\author{Limin Chen, Zhiwen Tang, Grace Hui Yang}
 \email{{lc1117,zt79,grace.yang}@georgetown.edu}
 \affiliation{%
   \institution{InfoSense, Department of Computer Science, Georgetown University}
  }

\begin{abstract}
    
Interactive Information Retrieval (IIR) and Reinforcement Learning (RL) share many commonalities, including an agent who learns while interacts, a long-term and complex goal, and an algorithm that explores and adapts. To successfully apply RL methods to IIR, one challenge is to obtain sufficient relevance labels to train the RL agents, which are infamously known as sample inefficient. However, in a text corpus annotated for a given query, it is not the relevant documents but the irrelevant documents that predominate.  
This would cause very unbalanced training experiences for the agent and prevent it from learning any policy that is effective. Our paper addresses this issue by using domain randomization to synthesize more relevant documents for the training. Our experimental results on the Text REtrieval Conference (TREC) Dynamic Domain (DD) 2017 Track show that the proposed method is able to boost an RL agent's learning effectiveness by 22\% in dealing with unseen situations.  

\end{abstract}



\maketitle

\section{Introduction}

Reinforcement Learning (RL) fits Interactive Information Retrieval (IIR) well as both of them center on accomplishing a goal during an interactive process. In RL, a machine agent maximizes its cumulative rewards collected during the course of interactions with an environment. In IIR, a search system satisfies an information need during the course of interactions with a user and a corpus. These commonalities have inspired approaches for IIR using RL solutions \cite{zhao2019deep,tang2019dynamic}. In these solutions, the search system is the RL agent; and both the user and text corpus the RL environment. 

Training an RL agent could be difficult. First, it is expensive to train the agent. To learn some policy useful, the agent may need to take millions of steps to interact with the environment. It becomes even more expensive when  real humans are involved in the interactions,  as what we have here in IIR. Simulations have therefore been proposed to replace real human users to train interactive systems~\cite{trecdd17}. 
Second, these simulators are usually created based on pre-annotated ground truth corpora. The corpora usually contain  many irrelevant documents and a few relevant documents. Therefore the simulators would not be able to form a balanced training environment. 



The unbalance in the forming of training environments would prevent the RL agent from learning good policies. It is because the rewards -- relevant documents -- are too few and the RL agent may not be able connect a long sequence of retrieval actions to a distant future reward, thus will never learn how to perform a task. This is known as the  problem of ``sparse rewards". Trained with sparse rewards, when the RL agent is deployed in a real dynamic environment, such as in the Web, it would be likely to make wrong decisions, especially when the agent encounters documents never seen before.



In this paper, we propose a novel domain randomization method to enhance the training experiences of RL agents in interactive retrieval.  Our method, {\it Document Environment Generation (DEG)},  proposes to automatically generate positive learning environments  as many as needed during a simulated training. DEG derives a stream of synthetic environments from available relevant documents by merging relevant segments within the documents with other irrelevant segments extracted from the corpus. 
In addition, our method  dynamically restrains the changing rate of a policy to make sure that the RL agent adapts conservatively to the synthetic environments.
We experiment our method on the Text REtrieval Conference (TREC) 2017 Dynamic Domain (DD) Track and the results show that the proposed method is able to statistically significantly boost the agents' learning effectiveness.



\section{Related Work}






Reinforcement learning has been successfully applied in several applications in text domains. They include dialogue systems \citep{li-etal-2016-deep}, recommendation systems \citep{liu2018deep} and dynamic search \citep{tang2019dynamic}. These systems aim to find items that match with user's interests by modeling the interactions between a system and a human user. The required participation of real human users makes it quite costly to sample training trajectories for the system. Researcher have proposed to build simulators to reduce the sample complexity \cite{trecdd17}. However, the difference between a simulation and a real environment, known as their reality gap, makes it challenging to directly deploy an RL agent into the real world~\cite{chebotar2019closing}. Domain randomization  has been proposed to alleviate this problem in robotics. For instance, \cite{tobin2017domain} randomized the objects and textures when training a robotic arm to grasp in cluttered environments. \cite{peng2018sim} randomized  the mass and damping of  robotic arms.  \cite{sadeghi2016cad2rl}  randomized the floor plan when training a drone to fly indoor. 

A similar technique, data augmentation, has been used in supervised learning to add more training data. In computer vision, it is proposed to augment image data by performing geometry transformation, such as flipping and cropping \citep{shorten2019survey}, and photometric transformation, such as edge enhancement and color jittering \citep{taylor2017improving}. In natural language processing,  it is proposed to generate new text by various language generation methods. For instances, \cite{li2017robust} created noisy texts by substituting words with its synonyms. \cite{ebrahimi-etal-2018-hotflip} flipped the order of characters to improve the robustness of a text classifier. \cite{wei2019eda} also augmented training samples by random insertion and deletion.  

In this work, we randomize a process to generate new texts by leveraging a unique characteristic in IR.  The characteristic is that whether a document is relevant largely depends on if it contains the query keywords and only the matching parts in the document would determine the relevance of the document. 

Training an RL agent in a series of different environments sometimes would result in ``catastrophic forgetting" \cite{kirkpatrick2017overcoming}. We handle this problem in our work too because our agent needs to learn from an enlarged and more diverse set of environments. \cite{rusu2016progressive} dealt with this problem by training an ensemble of neural networks, each for a separate task, and sharing weights among them.  \cite{kirkpatrick2017overcoming} augmented a loss function to  slow down the learning on important parameters. \cite{berseth2018progressive} transferred policies learned among different tasks with network distillation. Unlike them, DEG handles this problem by restraining the policy's changing rate based on how much  the newly generated environments differ from  the original.

\begin{figure*}[ht]
    \centering
    \includegraphics[width=0.73\textwidth]{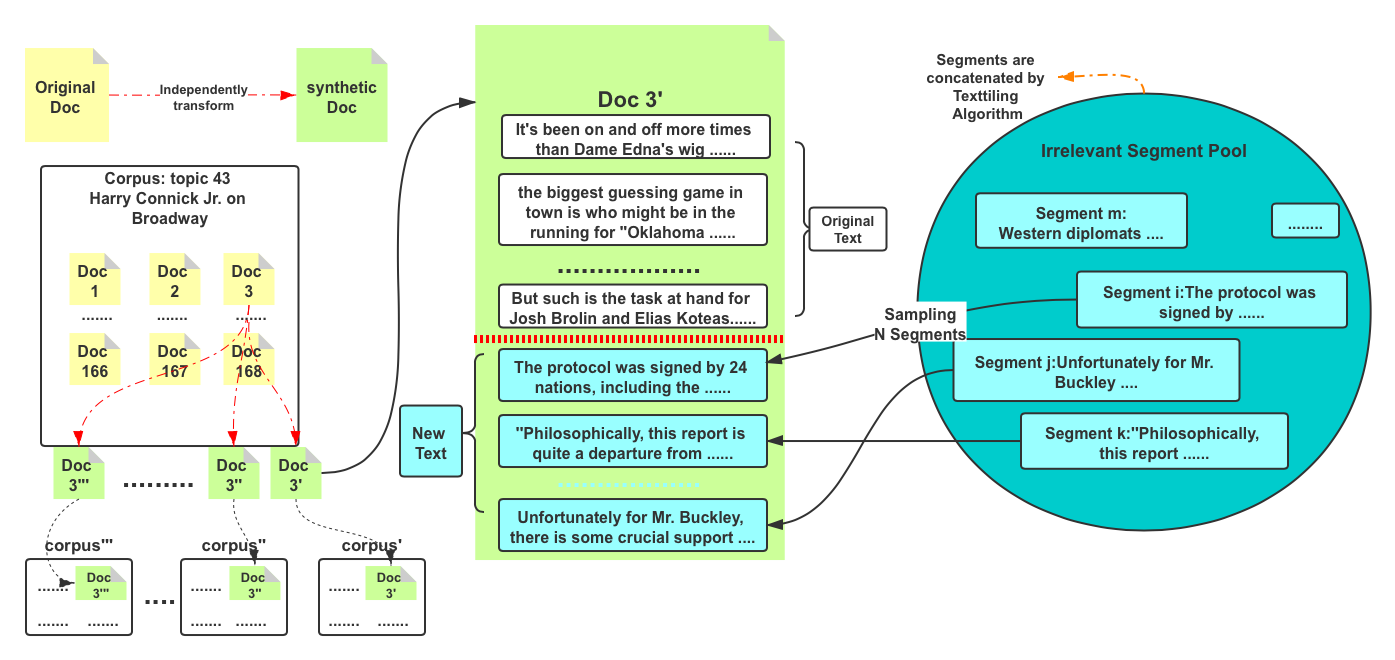}
    \caption{
    Generating new relevant documents $Doc 3'$, $Doc 3''$ and $Doc 3'''$ independently from $Doc 3$ (Topic DD17-43). 
    } 
    \label{fig:generate_doc}
\end{figure*}

\section{Setup}






In IIR, a human user searches for relevant information within a text collection from an interactive search engine.  The Text REtrieval Conference (TREC) Dynamic Domain (DD) Track \cite{trecdd17} provides a platform to evaluate interactive search engines. In the Track, a simulated user starts an initial query. At each subsequent time step, a search system  retrieves a set of documents and returns the top $K$ ($K=5$ in TREC DD) to the simulated user. The simulated user provides explicit feedback regarding how relevant these documents are and then the search system  adjusts its search algorithm based on the feedback.  The process repeats until the search stops. 

CE3 \cite{ce3}  is a state-of-the-art IIR system. It is based on the proximal policy optimization (PPO) algorithm~\cite{PPO}. 
In this paper, we improve this state-of-the-art system by incorporating domain randomization. 
Following CE3's practice, DEG also uses a corpus-level state representation. To build the states, our method first segments each document into a fixed number ($M$) of segments. Each segment  is on the same topic and maps into a separate doc2vec vector after compression. State at time $t$ is formed like taking a snapshot of the entire corpus by stacking together the representations of all documents at $t$. This global state representation at $t$ is expressed in the embedding function $\mathcal{S}$: 
\begin{equation}\label{eq:state_defn}
    s_t = \mathcal{S} (C, \mathcal{D}_1 \cup \mathcal{D}_2 ... \cup \mathcal{D}_{t-1})
\end{equation}
where $C$ is the text corpus and $\mathcal{D}_i$ is the set of documents retrieved at time $i$.  
The state representation's dimension is $|C| \times M \times n$, where $|C|$ is the size of the corpus, $M$ the number of segments per document, and $n$ the lower dimension ($n <<$ vocabulary size) of the doc2vec vector after compression. 
The retrieved documents at time $t$, i.e. $\mathcal{D}_t$, will be marked as `visited' on this global representation and passed on to next run of retrieval. 


Action at time $t$, $a_t$, is a weighting vector for document ranking. The ranking function calculates a relevance score between document $d_i$ and search topic (query) $q$ as the weighted sum over all segments $x_{ij} \in d_i$:
\begin{equation}\label{eq:action_defn}
\resizebox{0.5\linewidth}{!}{
    ${score}_{i,t} = f(a_t, d_i) =\sum_{j=1}^{M} x_{ij} \cdot a_t$
}
\end{equation} 
where 
$x_{ij}$ is the doc2vec representation of the $j^{th}$ segment in the $i^{th}$ document and $a_t$ is the action vector (weighting vector) at time $t$.

Reward $r_t$ is derived from the relevance ratings annotated by  human assessors. These relevance ratings are from 0 (irrelevant) to 4 (highly relevant). We define $r_t$ as the sum of all relevance ratings for the returned documents $D_t$ after duplicated results are removed:  
\begin{equation}\label{eq:reward}
r_t = \sum_{d_i \in \mathcal{D}_t \backslash (\mathcal{D}_1 \cup \mathcal{D}_2 \cup ... \cup \mathcal{D}_{t-1}) }  rel(d_i)
\end{equation}
where $rel()$ gets the relevance rating for  $d_i$ by looking up and summing the ratings for all passages in $d_i$ in the ground truth. 





\section{Proposed Method}

This section presents our method on generating relevant documents to form more balanced training environments and using an adaptive clipping rate to constrain the policy from  dramatic changes. 

\subsection{Generate Synthetic Environments}

Information retrieval is a task driven by keyword matching. Users recognize relevant documents by recognizing query keywords that are present in the documents. As long as the keywords are kept in a document, the document is considered as relevant. We therefore propose to separate relevant segments from irrelevant segments in a relevant document and put them into different uses.



Our process to create new relevant documents is the following. First, DEG separates the corpus into three parts, relevant segments from relevant documents, irrelevant segments from relevant documents, and irrelevant segments from irrelevant documents. Note all segments in an irrelevant document are irrelevant. Second, an `irrelevant pool" of segments, $P$, is formed by putting together  all irrelevant segments from both relevant and irrelevant documents. Third, for each relevant document, DEG samples $N$ segments, $p_1, p_2, ..., p_N$, from the irrelevant pool $P$:
\begin {equation}
N \sim \mathcal{U}(0, F)
\end{equation}
where $\mathcal{U}$ is a uniform distribution and $F$ is a parameter to control the maximum ``fake'' level that the synthesized documents would be.  Fourth, irrelevant segments $p_1, p_2, ... , p_N$ replace the original irrelevant segments in the relevant document and form a synthetic relevant document. Each sampling can independently generate a new document from the same seed document. Fifth, documents synthesized from the same document share the same relevance rating as the seed document as as they all contain the same set of relevant segments. 
Last, the process repeats for every relevant document in the corpus. DEG can generate as many new relevant documents as possible from the same original document. Figure \ref{fig:generate_doc} illustrates DEG's document generation process.


These generated documents are added into the corpus to form a new training environment. With more relevant documents now in the training environment, the RL agent is able to meet a reward (relevant document) more frequently when it interacts with the environment. It is then able to learn a policy with enough reward signals. Moreover, the agent can  be exposed to many different training environments when we include different synthesized documents to form a new environment. The agent is thus trained with not only more data but more diverse data, which would help with its generalizing ability. 

Note that the generation process needs knowledge about passage-level ground truth relevance, which may not be available in datasets outside of TREC DD. However, such knowledge may also be obtained from matching keywords with ground truth documents. The relevance can also be derived from implicit feedback such as clicks and dwell time. We think the method proposed here is applicable to other settings for interactive retrieval. The test environments used in this work do not use DEG because it would be unfair to know ground truth in a test environment. 

\subsection{Adaptive Training} 





 Training an RL agent on vastly different environments may result in a problem known as ``catastrophic forgetting" of the original environment \cite{kirkpatrick2017overcoming}. In our case, the synthetic environment could be very different from the original and causes this problem. We propose to control the policy derivations within a bound based on how different the synthesized environment is from the original. 
 We call this strategy ``adaptive training".  

The RL framework we use is based on PPO~\cite{PPO,ce3}, which optimizes the following objective: 
$\label{eq:Controlled Clipped Surrogate Objective}
J(\theta) = \hat{E_t}[min(r_t(\theta) \hat{A_t},clip(r_t(\theta),1-\epsilon,1+\epsilon)\hat{A_t})]
$, 
where 
 $r_t(\theta)=\frac{\pi_{\theta}(a_t|s_t)}{\pi_{old}(a_t|s_t)}$ is a ratio denoting the change of action distribution. The $clip()$ function limits the change ratio within $[1-\epsilon, 1+\epsilon]$, where $\epsilon$ is the maximum change of the action distributions and usually set to a fixed value, such as 0.2. This formulation has already been able to prevent drastic policy change. 
 
 Our work goes one step further to use a dynamically defined $\epsilon$, instead of a fixed value. In DEG, the synthetic environment differs from the original environment in terms of the documents it consists of. Thus, we calculate $\epsilon$ based on how much  the synthetic documents differ from the original document.  Their difference is measured by  Euclidean distance in the embedding space. For the $m^{th}$ synthetic environment,  $\epsilon_m$ is calculated as 
 \begin{equation} \label{eq:adaptive_epsilon}
     \epsilon_m = \alpha u^{-||\mathcal{S}(C) - \mathcal{S}(C'_m)||}
 \end{equation} 
 where $C$ is the original environment, $C'$ is the synthetic environment, $\mathcal{S}$ is the embedding function shown in Eq. \ref{eq:state_defn}, $||.||$ is the Euclidean distance function,  and $\alpha$ and $u$ are hyper parameters.



 For environments that differ more from the original environment,  $\epsilon_m$ puts a tighter restraint on the change rate, which  slows down the change of the policy. In this way, DEG prevents the agent from catastrophic forgetting when trained with  diverse environments. 

\section{Experiments}

We experiment on the TREC 
2017 Dynamic Domain Track \cite{trecdd17}. It uses the New York Times corpus \citep{sandhaus2008new}. 
60 search topics were created for the Track and each is a distinct learning environment.
We compare the performance of the following IIR methods: {\bf CE3} \cite{ce3}, a PPO-based RL method without domain randomization. 
{\bf DEG}, the proposed method with domain randomization and adaptive training;  
$\alpha=0.2$, $u=1.006$, and $F=5$.
{\bf DEG (fixed)}, a variant of DEG with a fixed $\epsilon$;  $\epsilon$ is set to $0.2$ as it works the best.

\subsection{Effectiveness under Unseen Situations}

\begin{figure}[t]
    \centering
    \includegraphics[width=0.45\columnwidth]{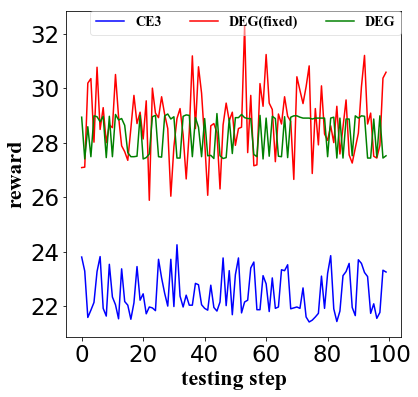}
    \caption{Average Reward}
    \label{fig:mean_plot_test}
\end{figure}

To understand how well a trained agent can perform interactive retrieval under unseen situations, we split the TREC DD 2017 dataset into non-overlapping training and test sets. The test dataset is considered as unseen environments to the agents.
For each search topic, we train RL agents using CE3, DEG or DEG (fixed) on the training set and test them on the test set.  We compare the agents' performance by measuring the rewards they obtain in the first 100 runs of interactions in the test set.  
Figure \ref{fig:mean_plot_test} reports the reward of CE3, DEG, and DEG (fixed), averaged over test search topics.

We observe that both DEG runs achieve a significant 5\% absolute  and 22\% relative ($p$<$0.0001$, double-tailed t-test) improvement over CE3. 
It shows that agents trained with domain randomization work much better than trained without it. 
We think it is because DEG generates a lot more relevant documents to form more balanced training environments and makes the agents generalize better. 
In addition, we found that although on average DEG and DEG (fixed) work comparable, DEG demonstrates much lower variance in the rewards that it can obtain. 
The proposed use of adaptive training contributes to a   more conservative policy update, which results in  more stable rewards. This is because DEG is less likely to take a big step to fall into a trajectory that is either too good or too bad.

\begin{figure}
    \centering
    \includegraphics[width=.21\textwidth]{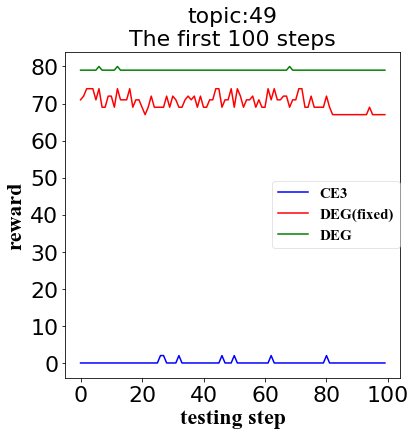}
    \includegraphics[width=.21\textwidth]{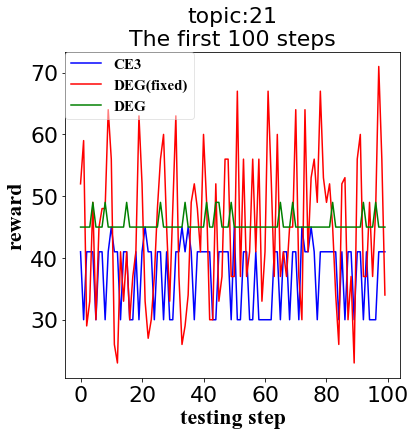}
    \caption{Case study: Topics 49 and 21} 
    \label{fig:testing_plot}
\end{figure}


\subsection{Case Studies}

Figure \ref{fig:testing_plot} shows two representative examples, topics 49 and 21. 
In Topic 49,  CE3 gets nearly zero rewards in unseen environments, which indeed works very poor. On the other hand, DEG variants  much better adjust to unseen environments and can both get impressive rewards. Domain randomization help the agents develop much better generalization ability.     
In Topic 21, we can see that DEG (fixed)'s rewards fluctuate a lot as the interactions go on while DEG produces much more stable rewards. Although DEG may miss some high-value steps, it too avoids the wrong moves, which is desirable for a task that cares much about precision at top positions.

\section{Conclusion}

    
 This paper proposes a domain randomization method to enhance  reinforcement learning  in interactive retrieval. Our method generates new relevant documents from existing documents in a collection to increase frequency of meeting relevant documents during RL training. The experiments show that this strategy can significantly boost an RL agent's performance in interactive retrieval under unseen situations. The proposed adaptive training method also helps the RL agent explore more steadily with stable rewards. 
 


\section*{Acknowledgements}
This research was supported by U.S. National Science Foundation Grant no. IIS-1453721. Any opinions, findings, conclusions, or recommendations expressed in this paper are of the authors, and do not necessarily reflect those of the sponsor.

\bibliographystyle{ACM-Reference-Format}

\bibliography{reference_short}
\end{document}